\documentclass[applsci,review,accept,moreauthors,pdftex,10pt,a4paper]{mdpi}

\firstpage{1}
\pubvolume{9}
\issuenum{6}
\articlenumber{1169}
\pubyear{2019}
\copyrightyear{2019}

\setitemize{parsep=6pt,itemsep=0pt,leftmargin=*,labelsep=5.5mm}
\setenumerate{parsep=6pt,itemsep=0pt,leftmargin=*,labelsep=5.5mm}
\setlist[description]{itemsep=0mm}

\history{Received: 27 November 2018; Accepted: 14 March 2019; Published: 19 March 2019}

\Title{Recent developments in the field of the metal-insulator transition in two dimensions}

\Author{Alexander~A. Shashkin$^{1}$ and Sergey~V.\ Kravchenko$^{2,}$*}

\AuthorNames{Alexander Shashkin and Sergey Kravchenko}

\address{%
$^{1}$ \quad Institute of Solid State Physics, Chernogolovka, Moscow District 142432, Russia; shashkin@issp.ac.ru\\
$^{2}$ \quad Physics Department, Northeastern University, Boston, Massachusetts 02115, USA; s.kravchenko@northeastern.edu}

\corres{Correspondence: s.kravchenko@northeastern.edu; Tel.: +1-617-413-4337}
\abstract{We review the latest developments in the field of the metal-insulator transition in strongly-correlated two-dimensional electron systems.  Particular attention is given to recent discoveries of a sliding quantum electron solid and interaction-induced spectrum flattening at the Fermi level in high-quality silicon-based structures.}

\keyword{Metal-insulator transition; strongly-correlated electrons; two-dimensional electron systems; low temperatures}
\begin{document}

\section{Introduction}

The zero-magnetic-field metal-insulator transition (MIT) was discovered in a strongly-interacting two-dimensional (2D) electron system in silicon metal-oxide-semiconductor field-effect transistors (MOSFETs) \cite{zavaritskaya1987,kravchenko1994,popovic1997} and later reported in a wide variety of 2D electron and hole systems: $p$-type SiGe heterostructures, $p$- and $n$-type GaAs/AlGaAs heterostructures, AlAs heterostructures, ZnO-related heterostructures, \textit{etc}.\ (for reviews, see Refs.~\cite{abrahams2001,kravchenko2004,shashkin2005,spivak2010}).  The strongest drop of the resistance on the metallic side of the transition (up to a factor of 12) at sub-Kelvin temperatures was reported in 2D systems in SiGe/Si/SiGe quantum wells \cite{melnikov2019}; in spite of lower disorder, the drop of the resistance in GaAs-based structures never exceeded a factor of about 3.  This discrepancy has been attributed particularly to the fact that electrons in silicon-based structures have two almost degenerate valleys, which further enhances the correlation effects \cite{punnoose2002,punnoose2005}.

Here we briefly review the experimental data obtained in different strongly correlated 2D electron and hole systems in zero and non-zero magnetic fields on both sides of the metal-insulator transition; the term strongly correlated means that the interaction parameter given by the ratio of the Coulomb and Fermi energies, $r_{\text{s}}=g_{\text{v}}/(\pi n_{\text{s}})^{1/2}a_{\text{B}}$, exceeds $r_{\text{s}}\sim10$ (here $g_{\text{v}}=2$ is the valley degeneracy, $n_{\text{s}}$ is the areal density of carriers, and $a_{\text{B}}$ is the effective Bohr radius in semiconductor).  Then we will discuss the behavior of the effective electron mass and the Land\'e $g$-factor on the metallic side of the transition.  Finally, transport evidence for the formation of a quantum electron solid on the insulating side of the transition will be reviewed.

\section{Metal-insulator transition in zero magnetic field}

The existence of a metallic state at zero magnetic field in 2D was first predicted by Finkelstein \cite{finkelstein1983,finkelstein1984} and Castellani~\textit{et al}.~\cite{castellani1984}.  The combined effects of interactions and disorder were studied by perturbative renormalization group methods.  The possibility of metallic behavior was confirmed later in Refs.~\cite{castellani1998,punnoose2002,punnoose2005}.

\begin{figure}
\centering
\includegraphics[width=4.5 cm]{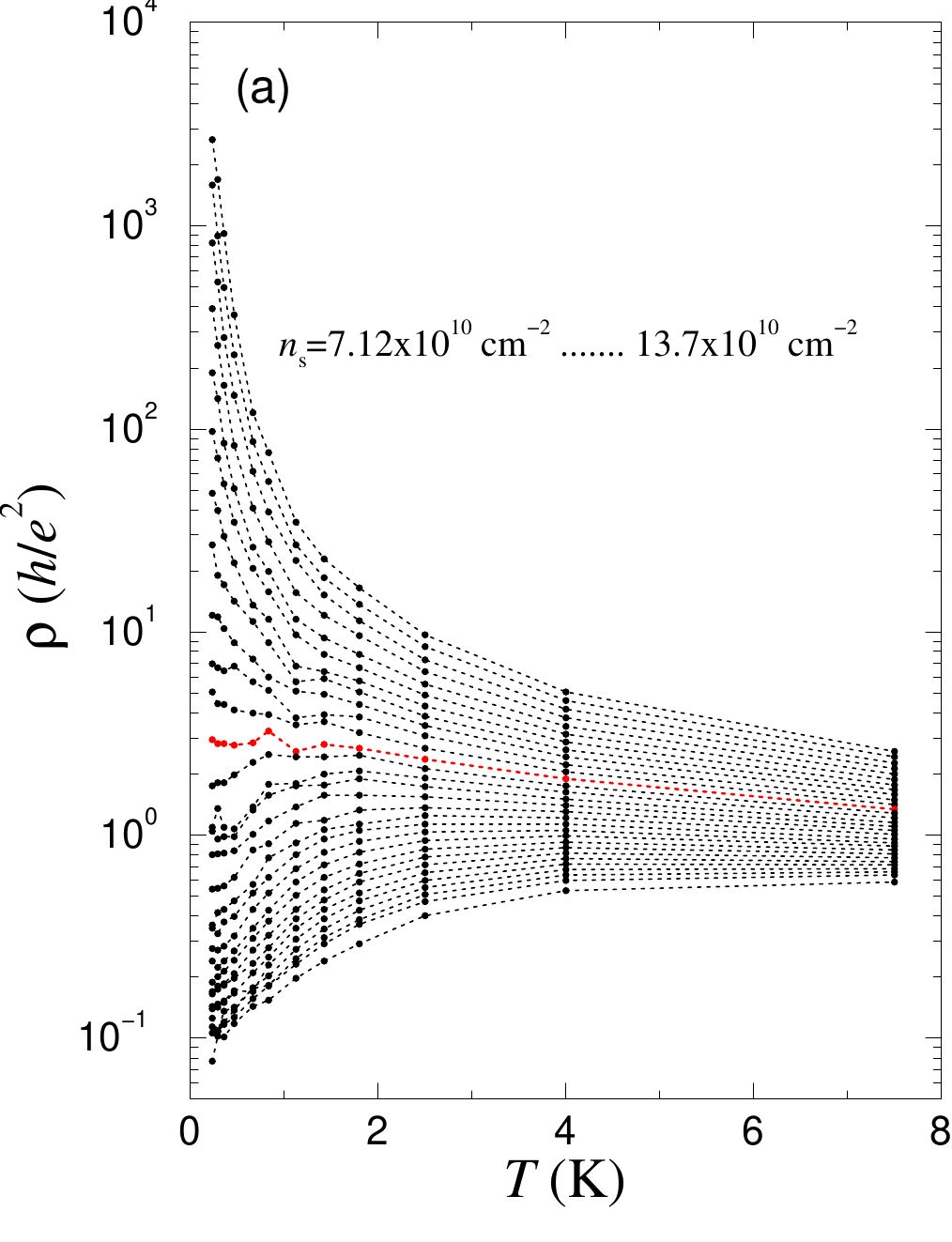} \includegraphics[width=5 cm]{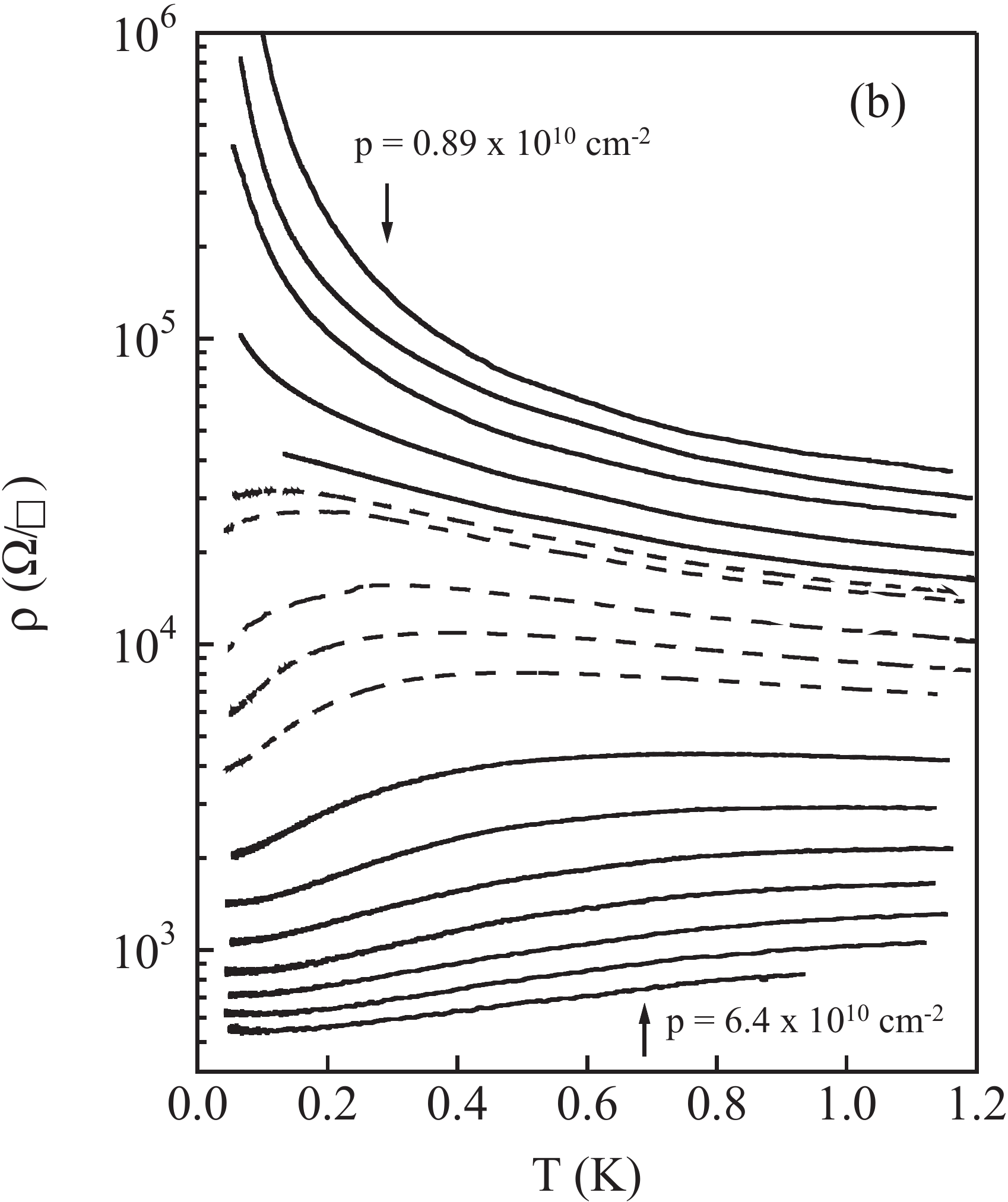} \includegraphics[width=5.5 cm]{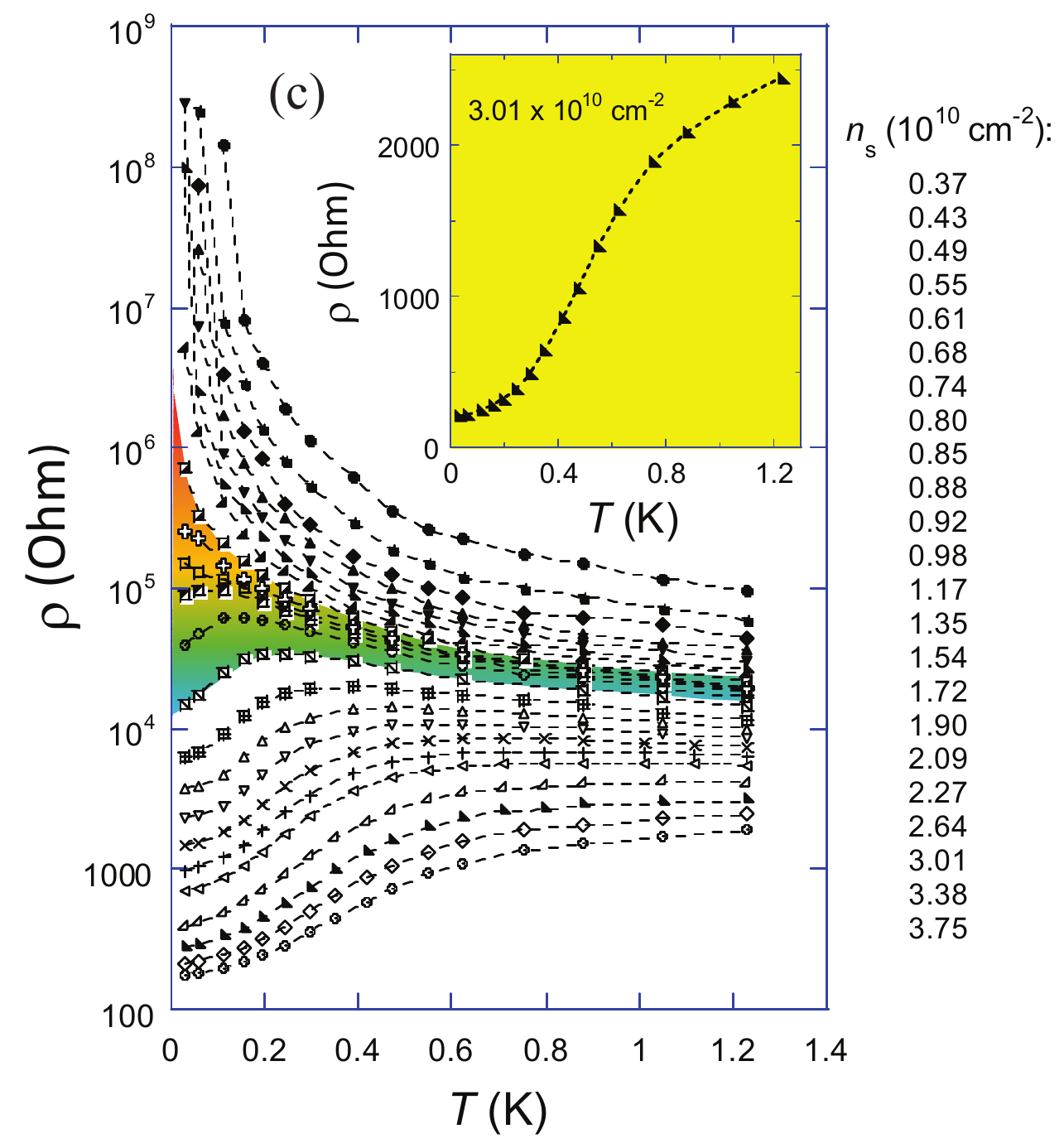}
\caption{Resistivity as a function of temperature in a Si MOSFET (\textbf{a}), $p$-type GaAs/AlGaAs heterostructure (\textbf{b}), and SiGe/Si/SiGe quantum well (\textbf{c}).  Figures adapted from \cite{kravchenko1995}, \cite{hanein1998}, and \cite{melnikov2019}, respectively.}\label{fig1}
\end{figure}

Typical experimental data showing the resistivity $\rho$ as a function of temperature $T$ for different electron densities in various systems are presented in Fig.~\ref{fig1} \cite{kravchenko1995,hanein1998,melnikov2019}. At lowest temperatures, the resistivity exhibits a strong metallic temperature dependence (${\rm d}\rho/{\rm d}T>0$) for electron densities above a well defined critical value, $n_{\text c}$, and insulating behavior (${\rm d}\rho/{\rm d}T<0$) for $n_{\text s} < n_{\text c}$. One refers to regime at $n_{\text s} > n_{\text c}$ as metallic and to that at $n_{\text s} < n_{\text c}$ as insulating \cite{shashkin2001}. The two regimes are separated by a separatrix, indicated in Fig.~\ref{fig1}~(a) by a dotted red line. This curve is often tilted in agreement with Ref.~\cite{punnoose2005} but becomes flat in the low-temperature limit or over an extended temperature range in the most uniform samples \cite{kravchenko1999}. The $\rho(T)$ dependences on the metallic side of the transition at $n_{\text s}$ just above $n_{\text c}$ are non-monotonic: while at temperatures exceeding a density-dependent value $T_{\text m}$, the derivative ${\rm d}\rho/{\rm d}T$ is negative, it changes sign at temperatures below $T_{\text m}$.

However, to determine the critical density for the MIT, one cannot rely on the derivative criterion alone.  In Ref.~\cite{shashkin2001}, additional method was used based on vanishing activation energy and vanishing nonlinearity of the current-voltage characteristics (looking from the insulating side of the transition).  It was shown that in clean Si MOSFETs in zero magnetic field, the critical density $n_{\text c}^*$, determined by such a method, coincides with the critical density $n_{\text c}$ determined by the derivative criterion.

\begin{figure}
\centering
\includegraphics[width=5 cm]{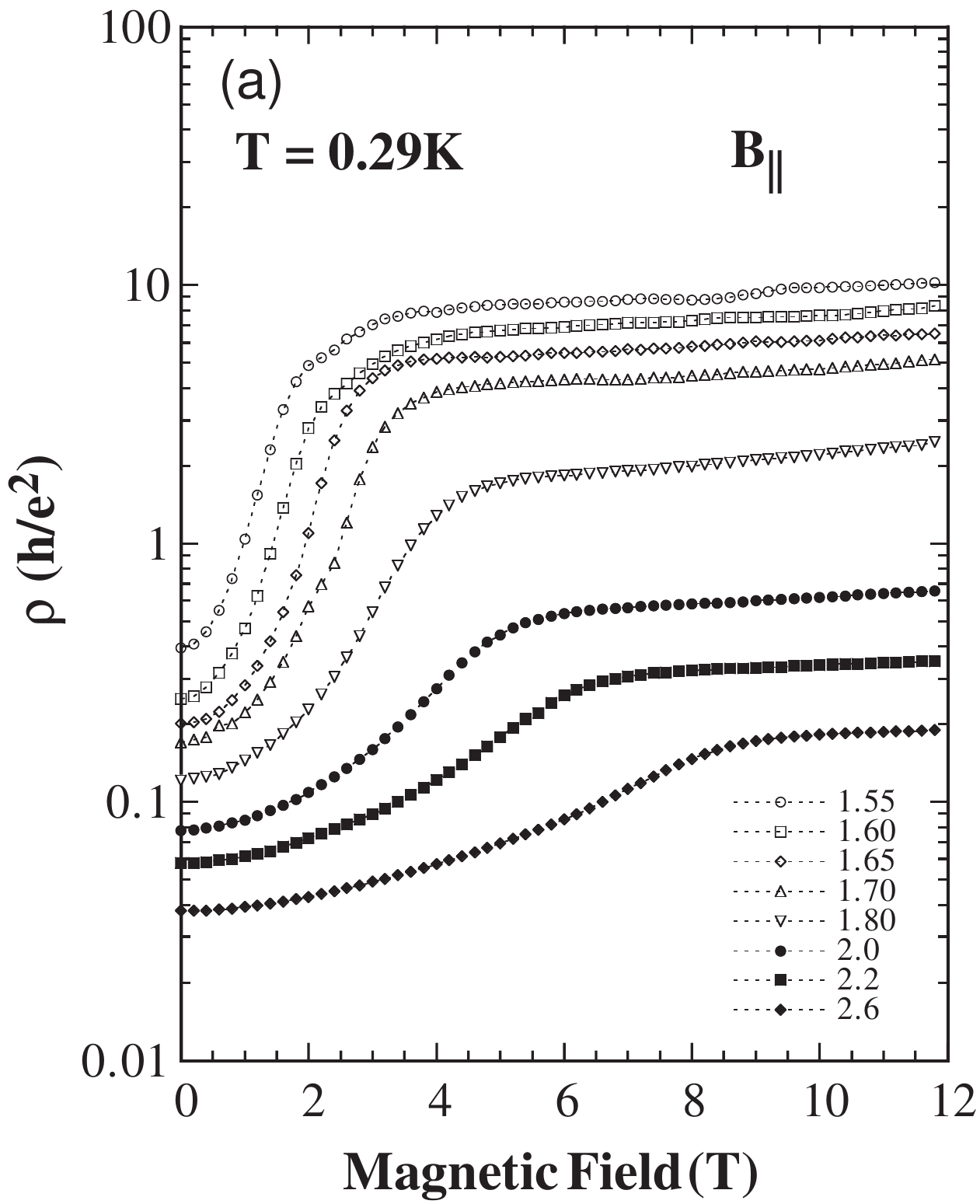} \includegraphics[width=5 cm]{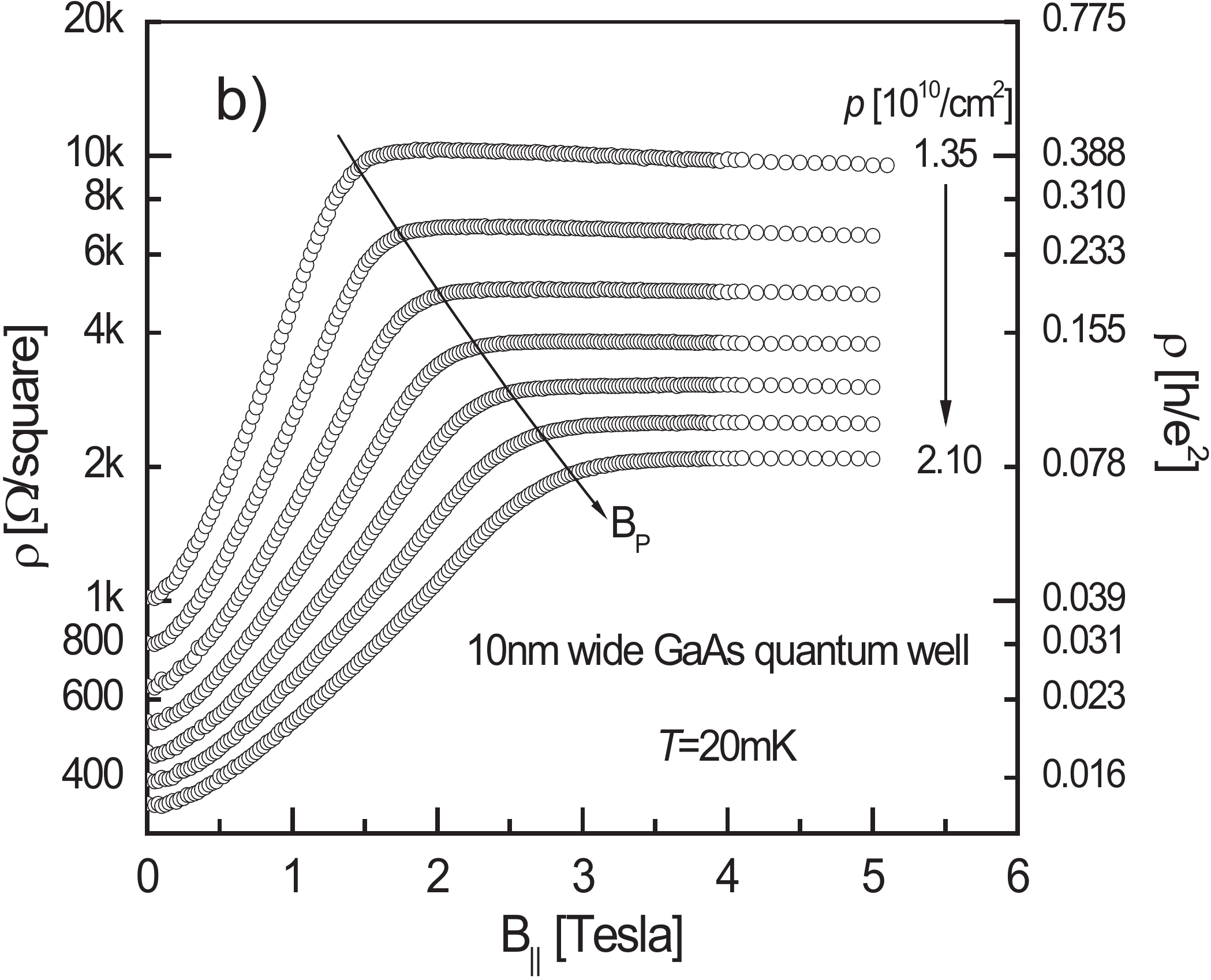}
\caption{(\textbf{a}) Resistivity vs.\ parallel magnetic field measured at $T=0.29$~K in a Si MOSFET. Densities are indicated in units of $10^{11}$~cm$^{-2}$; from Ref.~\cite{pudalov1997}. (\textbf{b}) Resistivity as a function of $B_{\|}$ in a 10~nm wide $p$-type GaAs quantum well at 20~mK; from Ref.~\cite{gao2006}.}\label{fig2}
\end{figure}

\section{Influence of the magnetic field parallel to the 2D plane}

The magnetoresistance $\rho(B_{\|})$ in magnetic fields $B_{\|}$ parallel to the 2D plane is strong and positive \cite{simonian1997,pudalov1997,gao2006} (examples are shown in Fig.~\ref{fig2}), but eventually saturates above some density-dependent magnetic field $B_{\|}>B^{*}$.  In sufficiently thin quantum wells, $B_{\|}$ has little effect on the orbital motion of the electrons and couples only to the electron spin; therefore, it is reasonable to attribute the saturation of the magnetoresistance to the complete spin polarization of the electron system \cite{dolgopolov2000}. Indeed, from the analysis of the positions of Shubnikov-de~Haas oscillations in tilted magnetic fields \cite{okamoto1999,vitkalov2000} it was concluded that the magnetic field $B^*$ is equal to that required to fully polarize the electrons' spins.

\begin{figure}
\centering
\includegraphics[width=9 cm]{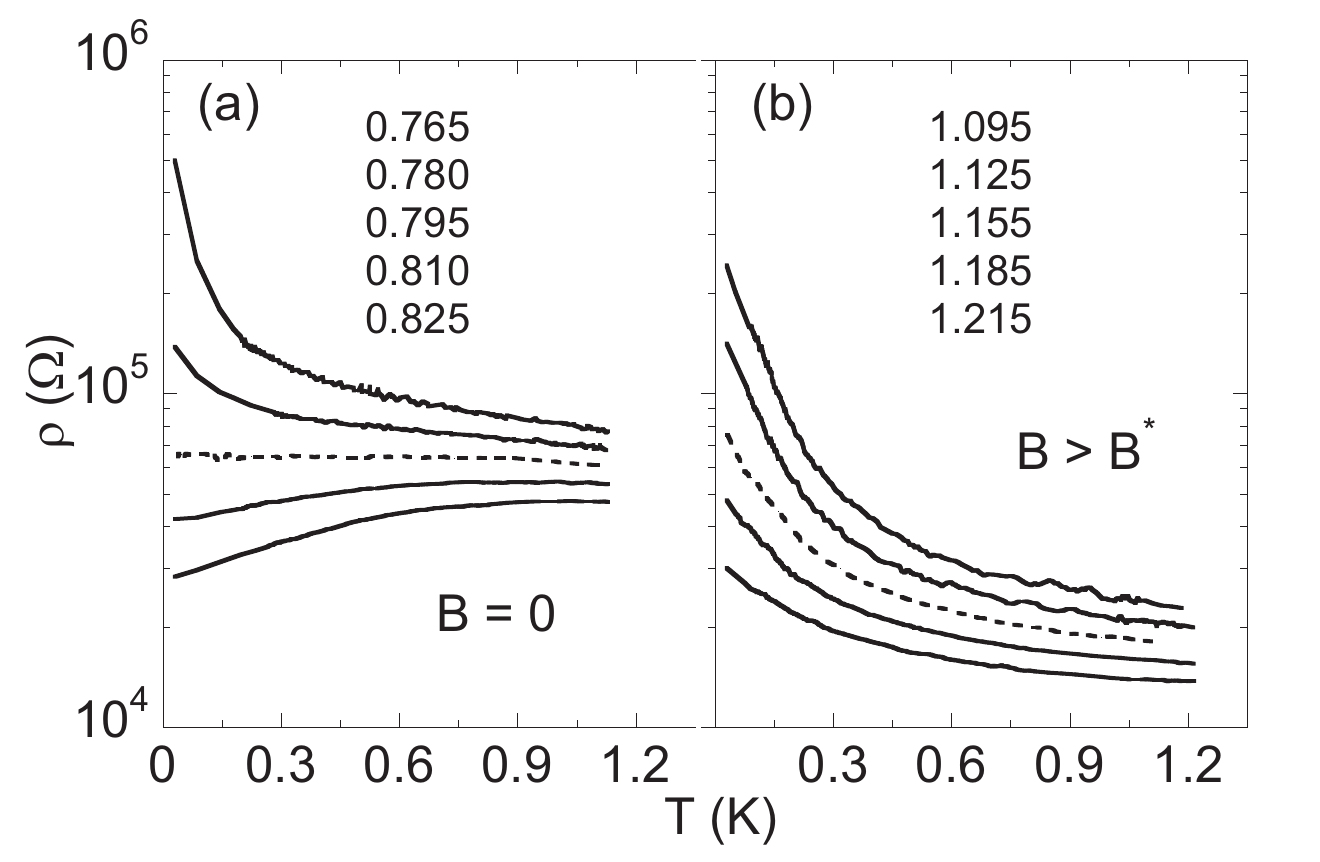}
\caption{Temperature dependence of the resistivity of a Si MOSFET at different electron densities near the MIT in zero magnetic field (a) and in a parallel magnetic field of 4 tesla (b). The electron densities are indicated in units of $10^{11}$~cm$^{-2}$. Dashed curves correspond to $n_{\text{s}} = n_{\text{c}}^*$ which is equal to $0.795\times 10^{11}$~cm$^{-2}$ in zero field and to $1.155\times 10^{11}$~cm$^{-2}$ in $B = 4$~tesla. Adapted from Ref.~\cite{shashkin2001}.}\label{fig3}
\end{figure}

Another effect of the parallel magnetic field is that it wipes out the strong metallic temperature dependence of the resistance at electron densities just above $n_{\text c}$ and suppresses the metallic regime \cite{spivak2010,dolgopolov1992,simonian1997,shashkin2001,dolgopolov2017,li2017}. In Fig.~\ref{fig3}, the temperature dependences of the resistivity in zero magnetic field (Fig.~\ref{fig3}~(a)) are compared with those measured in a parallel magnetic field, $B_{\|}=4$~tesla, high enough to cause full spin polarization (Fig.~\ref{fig3}~(b)). The dashed (middle) curves on both panels correspond to the critical electron densities, $n_{\text{c}}^*$, determined by the method of vanishing activation energy and vanishing nonlinearity of the current-voltage characteristics outlined above. The middle curve in Fig.~\ref{fig3}~(a), corresponding to $n_{\text{c}}^*=n_{\text{c}}$, is flat below 1~K. In sharp contrast with the $B = 0$ situation, at $B_{\|}>B^{*}$ not only are the $\rho(T)$ curves non-symmetric about the middle curve corresponding to $n_{\text{c}}^*$, but all of them have negative ``insulating-like'' derivatives ${\rm d}\rho/{\rm d}T<0$ in the entire temperature range, although the values of the resistivity are comparable to those in the $B = 0$ case. Moreover, in a strong parallel magnetic field, there is no temperature-independent $\rho(T)$ curve at any electron density \cite{shashkin2001}. Concerning the suppression of the metallic regime, it has been shown that the parallel magnetic field-induced increase in the critical electron density $n_{\text{c}}^*$ by the factor of about 1.4, observed in a strongly interacting two-dimensional electron system, is caused by the effects of exchange and correlations  \cite{dolgopolov2017}.

\begin{figure}
\center
\includegraphics[width=6.3 cm]{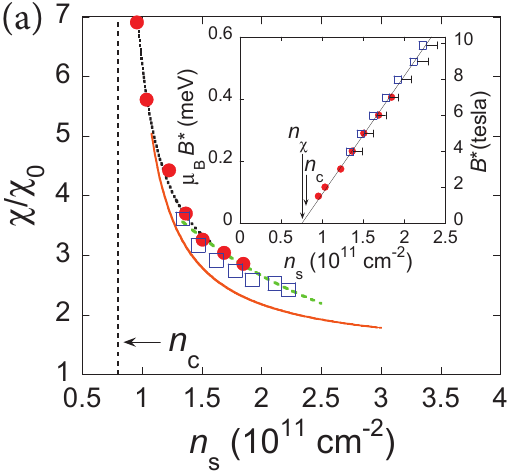}\includegraphics[width=7 cm]{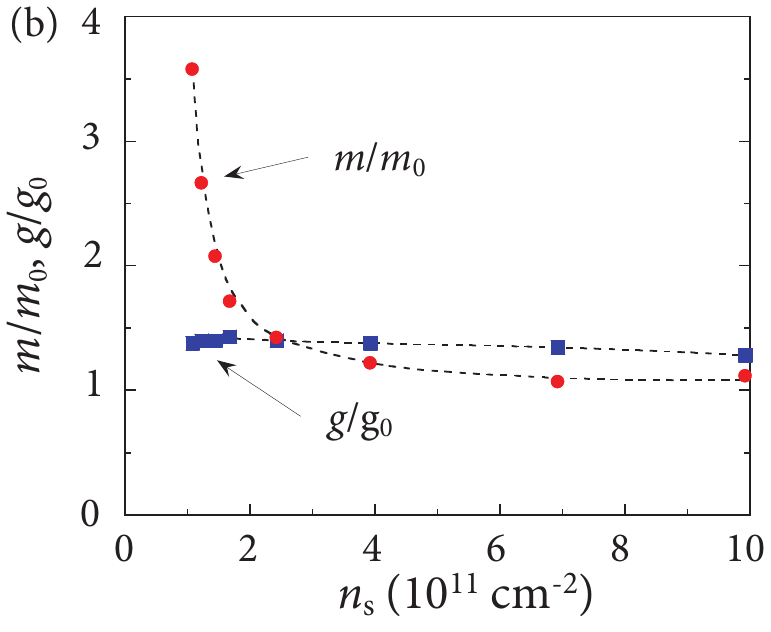}
\caption{(a)~Dependence of the Pauli spin susceptibility on electron density in a silicon MOSFET obtained by three independent methods: thermodynamic measurements of the magnetization (dashed line) and the magnetic field of the full spin polarization (circles), and magnetocapacitance (squares). The dotted line is a guide to the eye. Also shown by a solid line are the transport data of Ref.~\cite{shashkin2001a}. Inset: polarization field as a function of the electron density determined from the magnetization (circles) and magnetocapacitance (squares) data. The symbol size for the magnetization data reflects the experimental uncertainty. The dependence extrapolates linearly to zero field at a density $n_\chi$ just below $n_{\text c}$. Adapted from Ref.~\cite{shashkin2006}.
(b)~The effective electron mass $m$ (circles) and $g$-factor (squares) in a silicon MOSFET, determined from the analysis of the parallel field magnetoresistance and temperature-dependent conductivity, versus electron density. $m_0$ and $g_0$ are the electron mass and $g$-factor for non-interacting electrons in silicon.  The dashed lines are guides to the eye. Adapted from Ref.~\cite{shashkin2002}.}\label{fig4}
\end{figure}

\section{Spin susceptibility; \textit{g}-factor; the effective mass}

In Ref.~\cite{kravchenko2000} the Shubnikov-de~Haas oscillations in a 2D electron system in silicon were studied at low electron densities, and it was shown that near the metal-insulator transition, only ``spin'' minima of the resistance at Landau-level filling factors $\nu=$ 2, 6, 10, and 14 are seen, while the ``cyclotron'' minima at $\nu=$ 4, 8, and 12 disappear. A simple explanation of the observed behavior requires a giant enhancement of the ratio between the spin and cyclotron splittings near the metal-insulator transition.  Indeed, it turned out that with decreasing electron density (increasing interaction strength), the Pauli spin susceptibility, $\chi$, measured by several independent methods \cite{shashkin2001a,shashkin2006}, grows by as much as a factor of 7 compared to its non-interacting value, $\chi_0$, with a tendency to diverge at a disorder-independent density $n_{\chi}$ (see Fig.~\ref{fig4}~(a)). A similar increase of the spin susceptibility near the metal-insulator transition has been observed in single-crystalline ZnO-based heterostructures \cite{kozuka2014}.

In principle, this behavior can be due to either the increase of the effective electron mass or that of the Land\'e $g$-factor (or both).  In Ref.~\cite{shashkin2002}, these two values have been measured separately and it has been shown that the dramatic increase of the spin susceptibility is due to the strongly enhanced effective mass while the $g$-factor remains almost constant and close to its value in bulk silicon (see Fig.~\ref{fig4}~(b)).  The strong increase of the effective mass has been later independently confirmed by the analysis of the temperature dependence of Shubnikov-de~Haas oscillations \cite{shashkin2003} and magnetization measurements in perpendicular magnetic fields \cite{anissimova2006}.

In measurements of the thermopower in Si MOSFETs, data obtained for electron densities that are much closer to the critical point than the earlier measurements confirm that the critical density $n_\text{m}$ lies close to, but is consistently below the critical density $n_\text{c}$ for the MIT \cite{mokashi2012}. The results indicate the occurrence of an interaction-induced transition to a new phase at low density that may be a precursor phase or a direct transition to the Wigner solid.

It has been found recently that in contrast to previous experiments on lower-mobility samples, in ultra-high mobility SiGe/Si/SiGe quantum wells the critical electron density $n_\text{c}$ of the MIT becomes smaller than the density $n_\text{m}$, where the effective mass at the Fermi level tends to diverge \cite{melnikov2019}.

\section{Band flattening; possible condensation of fermions}

The creation and investigation of flat-band materials is a forefront area of modern condensed matter physics \cite{heikkila2011,ns2013,peotta2015,volovik15}. In particular, the interest is ignited by the fact that, due to the anomalous density of states, the flattening of the band may be important for the construction of room temperature superconductivity. The appearance of a flat band is theoretically predicted \cite{amusia14,camjayi08,yudin14} in a number of systems including high-temperature superconductors, heavy fermion systems, $^3$He, and 2D electron systems. As the strength of fermion-fermion interaction is increased, the single-particle spectrum becomes progressively flatter in the vicinity of the Fermi energy eventually forming a plateau. The flattening of the spectrum is related to the increase of the effective fermion mass $m_{\text F}$ at the Fermi level and the corresponding peak in the density of states.

The experimental data obtained in strongly correlated 2D electron systems can be divided into two groups:\\
(a)~data describing the electron system as a whole, like thermodynamic density of states, magnetization of the electron system, or the magnetic field required to fully polarize electron spins, and\\
(b)~data related only to the electrons at the Fermi level, like the amplitude of the Shubnikov-de~Haas oscillations.  This yields the effective mass $m_{\text F}$ and Land\'e $g$-factor $g_{\text F}$ at the Fermi level.

Usually, the data in the first group are interpreted using the language of quasiparticles, in which the energy-averaged effective mass, $m$, and Land\'e $g$-factor, $g$, are used. To determine the values, the formulas that hold for the case of non-interacting electrons are used. Although this approach is ideologically not correct, the results for $m$ and $g$ are often the same as, or close to, the results for $m_{\text F}$ and $g_{\text F}$. For example, simultaneous increase of the energy-averaged effective mass and that at the Fermi level in Si MOSFETs was reported in publications described in the previous section.

\begin{figure}[H]
\centering
\includegraphics[width=8 cm]{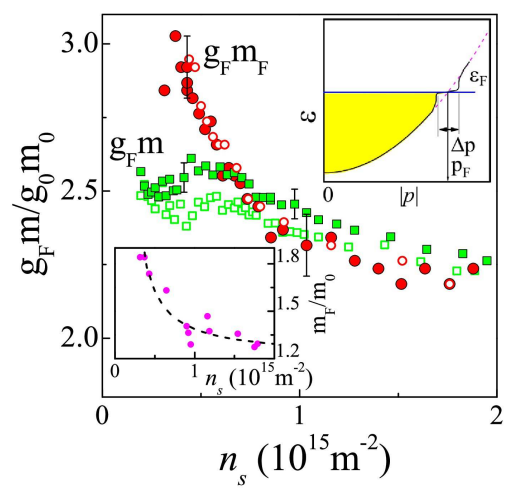}\vspace{-4cm}
\caption{Product of the Land\'e factor and effective mass as a function of electron density in a SiGe/Si/SiGe quantum well determined by measurements of the field of full spin polarization $B^*$ (squares) and Shubnikov-de Haas oscillations (circles) at $T\approx 30$~mK. The empty and filled symbols correspond to two samples. The experimental uncertainty corresponds to the data dispersion and is about 2\% for the squares and about 4\% for the circles. ($g_0=2$ and $m_0=0.19\, m_{\text e}$ are the values for non-interacting electrons, $m_{\text e}$ is the free electron mass, and $g_{\text F}\approx g_0$ is the $g$-factor at the Fermi level). The top inset shows schematically the single-particle spectrum of the electron system in a state preceding the band flattening at the Fermi level (solid black line). The dashed violet line corresponds to an ordinary parabolic spectrum. The occupied electron states at $T=0$ are indicated by the shaded area. Bottom inset: the effective mass $m_{\text F}$ versus electron density determined by analysis of the temperature dependence of the amplitude of Shubnikov-de Haas oscillations, similar to Ref.~\cite{melnikov2014}. The dashed line is a guide to the eye. From Ref.~\cite{melnikov2017}.}\label{fig5}
\end{figure}

In Ref.~\cite{melnikov2017}, the magnetic field required to fully polarize the electrons' spins in the ultra-high mobility 2D electron system in SiGe/Si/SiGe quantum wells of unprecedented quality \cite{melnikov2015} has been investigated. The product $g_{\text F}m$ that characterizes the whole 2D electron system can be determined in the clean limit from the equality of the Zeeman splitting and the Fermi energy of the spin-polarized electron system

\begin{equation}
g_{\text F}\mu_{\text B}B^*=\frac{2\pi\hbar^2n_{\text s}}{mg_{\text v}},\label{gm}
\end{equation}
where $\mu_{\text B}$ is the Bohr magneton. Simultaneously, the authors studied Shubnikov-de~Haas oscillations, yielding the product $g_{\text F}m_{\text F}$ at the Fermi level. It turned out (see Fig.~\ref{fig5}) that with decreasing electron density (or increasing interaction strength), the product $g_{\text F}m_{\text F}$ at the Fermi level monotonically increases in the entire range of electron densities, while the energy-averaged product $g_{\text F} m$ saturates at low densities. Since the exchange effects in the 2D electron system in silicon are negligible, this difference can only be attributed to the different behaviors of the two effective masses. Their qualitatively different behavior reveals a precursor to the interaction-induced single-particle spectrum flattening at the Fermi level in this electron system.

It seems natural to interpret these experimental results within the concept of the fermion condensation \cite{khodel1990,nozieres1992,zverev2012}.  The fermion condensation occurs at the Fermi level in a range of momenta, unlike the condensation of bosons. As the strength of electron-electron interactions increases, the single-particle spectrum flattens in a region $\Delta p$ near the Fermi momentum $p_F$ (top inset to Fig.~\ref{fig5}). At relatively high electron densities ($n_{\text s}>0.7\times 10^{15}$~m$^{-2}$), this effect is not important because the single-particle spectrum does not change noticeably in the interval $\Delta p$, and the behaviors of the energy-averaged effective mass and that at the Fermi level are similar. However, decreasing the electron density in the range $n_{\text s}<0.7\times 10^{15}$~m$^{-2}$ stimulates the flattening of the spectrum so that the effective mass at the Fermi level, $m_{\text F}=p_{\text F}/V_{\text F}$, continues to increase (here $V_{\text F}$ is the Fermi velocity). In contrast, the energy-averaged effective mass does not, being not particularly sensitive to this flattening.

\section{Transport evidence for a sliding quantum electron solid}

The insulating side of the metal-insulator transition has been extensively studied (see, \textit{e.g.}, Refs.~\cite{andrei1988,goldman1990,kravchenko1991,pudalov1993,knighton2018}), but no definitive conclusion has been reached concerning the origin of the low-density state. The observed nonlinear current-voltage ($I-V$) curves were interpreted as either manifestation of the depinning of an electron solid \cite{goldman1990,williams1991,kravchenko1991,pudalov1993,chitra2005} or the breakdown of the insulating phase within traditional scenarios such as strong electric field Efros-Shklovskii variable range hopping \cite{shklovskii1992} or percolation (see, \textit{e.g.}, Refs.~\cite{shashkin2005,jiang1991} and references therein). The observation of broad-band voltage noise at the threshold $I-V$ curves \cite{goldman1990,kravchenko1991,pudalov1993}, as well as the attempts to probe the low-density state in perpendicular magnetic fields \cite{diorio1990,diorio1992,qie2012,knighton2018,qie2018}, also have not yielded information that allows a choice between the depinning of the electron solid or traditional mechanisms \cite{shashkin2005,jiang1991}. Much confusion was introduced by the fact that many authors chose to interpret their data in terms of Wigner crystal \cite{wigner1934,chaplik1972,tanatar1989,attaccalite2002,dolgopolov2015}, ignoring mundane interpretations.

\begin{figure}
\centering
\includegraphics[width=8 cm]{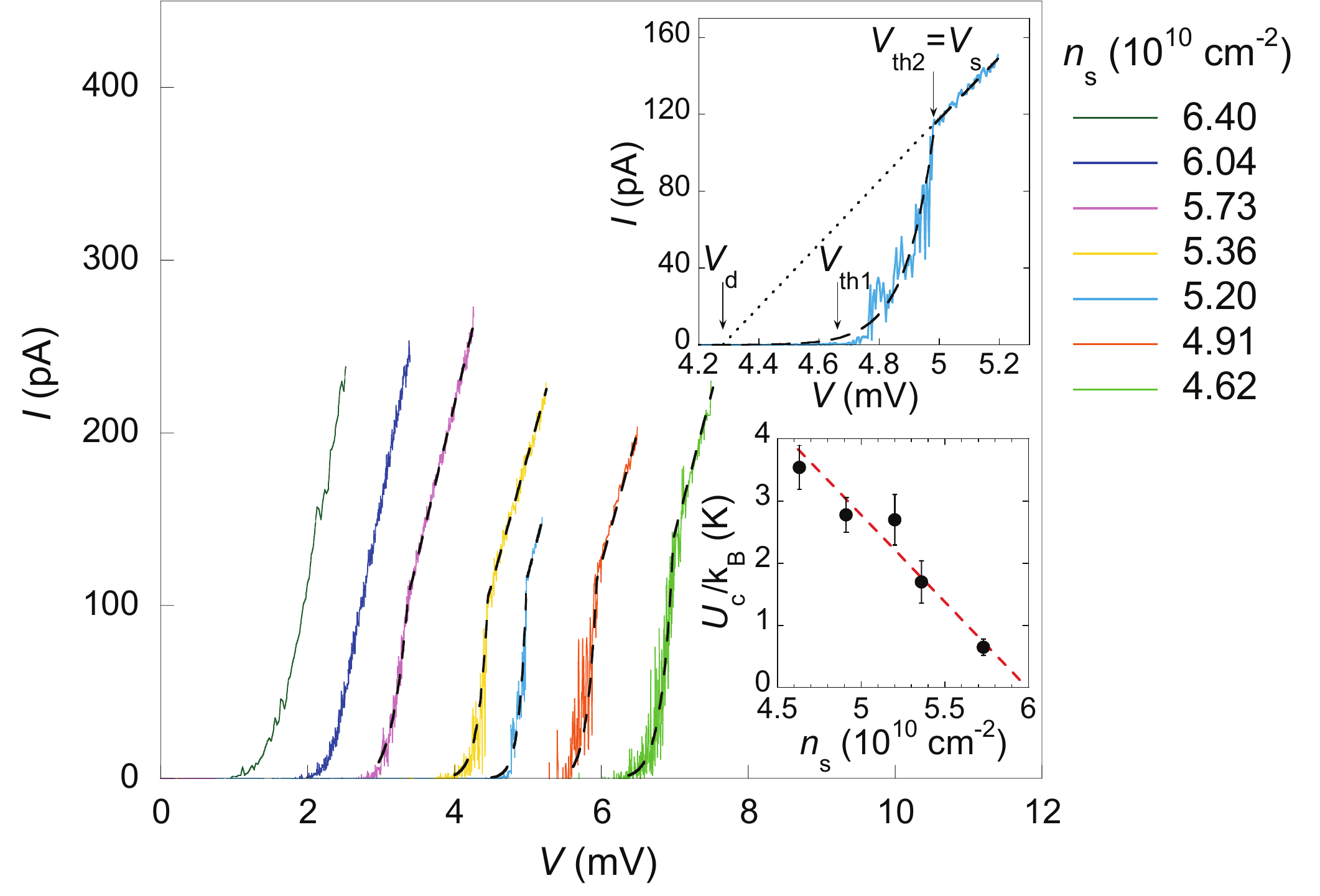}
\caption{$V-I$ curves are shown for different electron densities in the insulating state of a silicon MOSFET at a temperature of 60~mK. The dashed lines are fits to the data using Eq.~(\ref{I}). The top inset shows the $V-I$ curve for $n_{\text{s}}=5.20\times 10^{10}$~cm$^{-2}$ on an expanded scale; also shown are the threshold voltages $V_{\text{th1}}$ and $V_{\text{th2}}$, the static threshold $V_{\text{s}}=V_{\text{th2}}$, and the dynamic threshold $V_{\text{d}}$ that is obtained by the extrapolation of the linear region of the $V-I$ curve to zero current. Bottom inset: activation energy $U_{\text{c}}$ \textsl{vs}.\ electron density.  Vertical error bars represent standard deviations in the determination of $U_{\text{c}}$ from the fits to the data using Eq.~(\ref{I}).  The dashed line is a linear fit.  From Ref.~\cite{brussarski2018}.}\label{fig6}
\end{figure}

In Ref.~\cite{brussarski2018}, a significant breakthrough in the understanding of the origin of the low-density state has been reported. The authors observed two-threshold $V-I$ characteristics with a dramatic increase in noise between the two threshold voltages at the breakdown of the insulating state. The noise in current, in the form of fluctuations with time, dramatically increases above the first threshold, $V_{\text{th1}}$, and essentially disappears above the second threshold, $V_{\text{th2}}$. The double threshold behavior is very similar to that known for the collective depinning of the vortex lattice in Type-II superconductors (see, \textit{e.g.}, Ref.~\cite{blatter1994}) (provided the voltage and current axes are interchanged). This strongly favors the sliding 2D quantum electron solid whereas the double threshold behavior cannot be described within alternative scenarios (percolation or overheating). Rather than being an ideal Wigner crystal, however, the 2D electron system described here is likely to be closer to an amorphous solid, which is similar to the case of the vortex lattice in Type-II superconductors where the collective pinning was observed.

In figure~\ref{fig6}, several low-temperature voltage-current curves at different electron densities in the insulating regime $n_{\text{s}}<n_{\text{c}}$ are shown. The corresponding interaction parameter exceeds $r_{\text{s}}\sim20$. Two threshold voltages are observed at electron densities below $\approx 6\times 10^{10}$~cm$^{-2}$: with increasing applied voltage, the current stays near zero up to a first voltage threshold $V_{\text{th1}}$, then increases sharply until a second threshold voltage $V_{\text{th2}}$ is reached, above which the slope of the $V-I$ curve is significantly reduced and the $V-I$ curve is linear although not ohmic (see also the top inset to Fig.~\ref{fig6}). As the electron density is increased, the value of $V_{\text{th1}}$ decreases while the second threshold becomes less pronounced and eventually disappears. The authors observed no hysteresis in the range of electron densities studied. They point out that the observed behavior is quite distinct from the behavior reported in the insulating state in amorphous InO films, where the current was found to jump at the threshold voltage by as much as several orders of magnitude and the $V-I$ curves exhibited hysteresis consistent with bistability and electron overheating \cite{ovadia2009,altshuler2009}. It is also worth noting that the occurrence of the double threshold cannot be explained within the percolation picture in which case a single threshold is expected \cite{shashkin2005}. Thus, the double threshold behavior cannot be described within existing traditional models.

Two remarks are due here.  First, at electron density $n_{\text{s}}\approx6\times10^{10}$~cm$^{-2}$, below which the double-threshold behavior is evident, the interaction parameter is equal to $\approx22$.  This is below the critical value $r_{\text{s}}=37\pm5$ at which the quantum Wigner crystallization in a disorder-free 2D system is expected to happen in zero magnetic field, according to the numerical simulations \cite{tanatar1989}.  However, in subsequent simulations it was shown that the crystallization should occur at lower values of $r_{\text s}$ when disorder is present \cite{chui1995}.  Second, the disorder-independent divergence of the effective mass and the disorder-dependent formation of the quantum electron solid immediately imply the existence of an intermediate phase preceding the electron solidification.  Indeed, the existence of such a phase was predicted in Refs.~\cite{khodel1990,attaccalite2002,spivak2004}.

\begin{figure}
\centering
\includegraphics[width=7 cm]{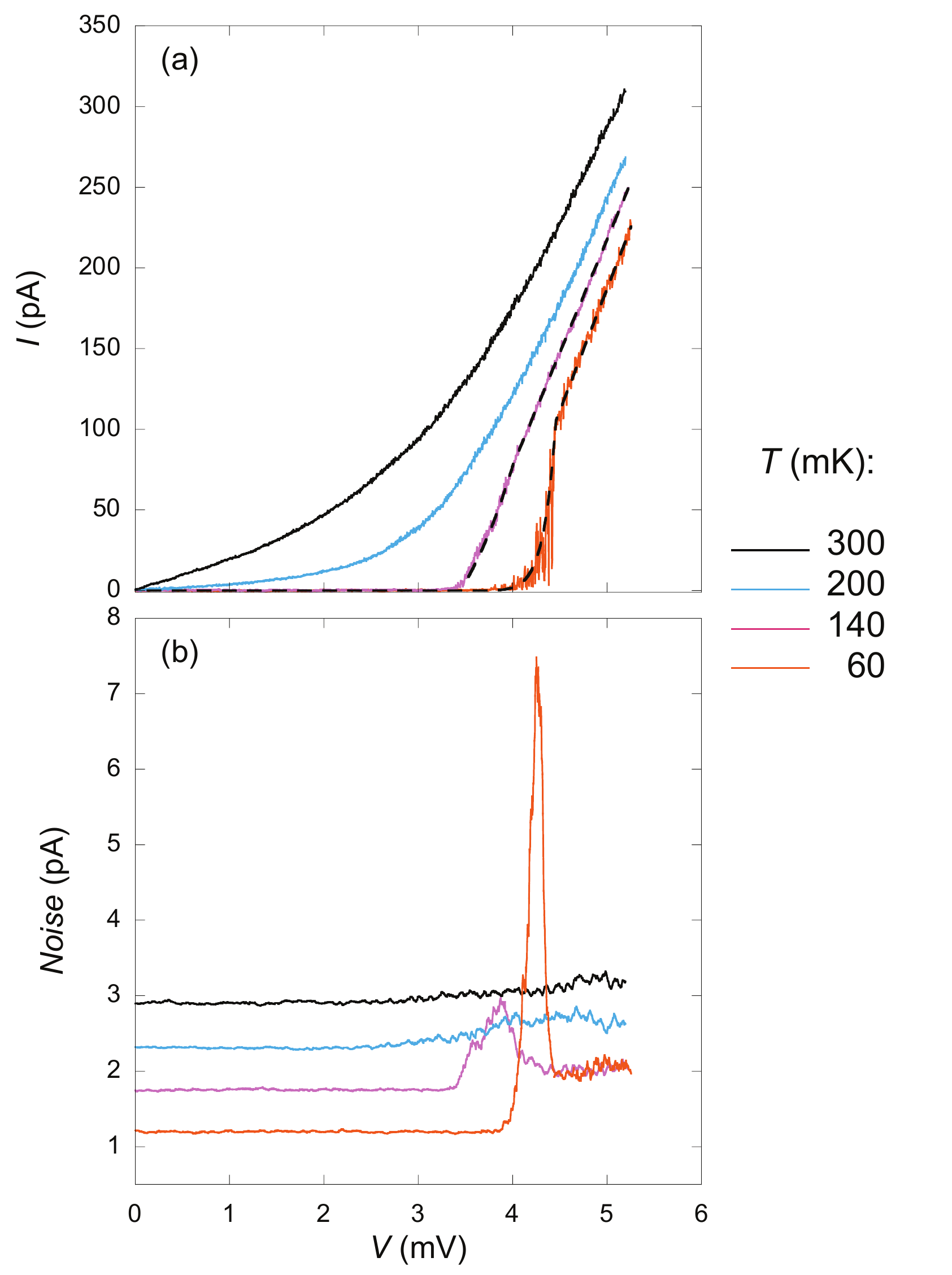}
\caption{(a) $V-I$ characteristics at $n_{\text{s}}=5.36\times 10^{10}$~cm$^{-2}$ in a silicon MOSFET for different temperatures. The dashed lines are fits to the data using Eq.~(\ref{I}). (b)~The broad-band noise as a function of voltage for the same electron density and temperatures.  The three upper curves are shifted vertically for clarity. From Ref.~\cite{brussarski2018}.}\label{fig7}
\end{figure}

Figure~\ref{fig7}~(a) shows the $V-I$ characteristics for $n_{\text{s}}=5.36\times 10^{10}$~cm$^{-2}$ at different temperatures. As the temperature, $T$, is increased, the second threshold $V_{\text{th2}}$ becomes less pronounced and the threshold behavior of the $V-I$ curves eventually smears out due to the shrinkage of the zero-current interval.  The measured broad-band noise is shown as a function of voltage in Fig.~\ref{fig7}~(b) for different temperatures at electron density $n_{\text{s}}=5.36\times 10^{10}$~cm$^{-2}$. A large increase in the noise is observed between the thresholds $V_{\text{th1}}$ and $V_{\text{th2}}$ at the lowest temperature. This large noise decreases rapidly with increasing temperature in agreement with the two-threshold behavior of the $V-I$ curves of Fig.~\ref{fig7}~(a). The frequency dependence of the noise amplitude is $1/f^{\alpha}$ law with $\alpha=0.6\pm0.1$, which is close to unity.

The authors of Ref.~\cite{brussarski2018} analyze their results in light of a phenomenological theory based on pinned elastic structures. Indeed, there is a striking similarity between the double-threshold $V-I$ dependences in the low-$n_{\text{s}}$ state of Si MOSFETs and those (with the voltage and current axes interchanged) known for the collective depinning of the vortex lattice in Type-II superconductors (see, \textit{e.g.}, Ref.~\cite{blatter1994}). The physics of the vortex lattice in the latter system, in which the existence of two thresholds is well documented, can be adapted for the case of an electron solid. In a superconductor, current flows for zero voltage; the depinning of the vortex lattice occurs when a non-zero voltage appears. In the authors' case, the situation is reciprocal: a finite voltage is applied, but at first the current is not flowing in the limit of zero temperature; the depinning of the electron solid is indicated by the appearance of a non-zero current. The intermediate regime between the dynamic ($V_{\text{d}}$) and static ($V_{\text{s}}$) thresholds corresponds to the collective pinning of the solid. In this region the pinning occurs at the centers with different energies and the current is thermally activated:

\begin{equation}
I\propto\exp\left[-\frac{U(V)}{k_{\text{B}}T}\right],
\label{exp}
\end{equation}
where $U(V)$ is the activation energy. The static threshold $V_{\text{s}}=V_{\text{th2}}$ signals the onset of the regime of solid motion with friction. This corresponds to the condition

\begin{equation}
eEL=U_{\text{c}}.
\label{Uc}
\end{equation}
(here $E$ is the electric field and $L$ is the characteristic distance between the pinning centers with maximal activation energy $U_{\text{c}}$). The balance of the electric, pinning, and friction forces in the regime of solid motion with friction yields a linear $V-I$ characteristic that is offset by the threshold $V_{\text{d}}$ corresponding to the pinning force

\begin{equation}
I=\sigma_0(V-V_{\text{d}}),
\label{linear}
\end{equation}
where $\sigma_0$ is a coefficient. (Near $V_{\text{d}}$, one can in general expect a power-law behavior of $(V-V_{\text{d}}$)). Assuming that the activation energy for the electron solid is equal to

\begin{equation}
U(V)=U_{\text{c}}-eEL=U_{\text{c}}(1-V/V_{\text{s}}),
\label{U}
\end{equation}
the authors obtain the expression for the current

\begin{equation}
I=\left\{\begin{array}{c}
\sigma_0(V-V_{\text{d}})\; {\text{ if }} V>V_{\text{s}}\\
\sigma_0(V-V_{\text{d}})\exp\left[-\frac{U_{\text{c}}(1-V/V_{\text{s}})}{k_{\text{B}}T}\right]\; {\text{ if }} V_{\text{d}}<V\leq V_{\text{s}}.
\end{array}\right.\label{I}
\end{equation}
The fits to the data using Eq.~(\ref{I}) are shown by dashed lines in Figs.~\ref{fig6} and \ref{fig7}~(a). Eq.~(\ref{I}) successfully describes the experimentally observed two-threshold $V-I$ characteristics. The value of $U_{\text{c}}$ decreases approximately linearly with electron density and reaches zero at $n_{\text{s}}\approx 6\times 10^{10}$~cm$^{-2}$ (the bottom inset to Fig.~\ref{fig6}). (Note that this is in contrast to the vanishing activation energy of electron-hole pairs at $n_{\text{c}}$ obtained by measurements of the resistance in the limit of zero voltages/currents \cite{shashkin2001}). Probably, the vanishing $U_{\text{c}}$ is related to the minimum number of the strong pinning centers for which the collective pinning is still possible. The coefficient $\sigma_0$ is approximately constant ($\sigma_0\approx 1.6\times 10^{-7}$~Ohm$^{-1}$), and this indicates that the solid motion with friction is controlled by weak pinning centers \cite{blatter1994}. The authors of Ref.~\cite{brussarski2018} argue that the large noise in the regime of the collective pinning of the solid between $V_{\text{d}}$ and $V_{\text{s}}$ should be suppressed in the regime of solid motion with friction at $V>V_{\text{s}}$. Indeed, in the regime of the collective pinning of the solid between $V_{\text{d}}$ and $V_{\text{s}}$, the solid deforms locally when the depinning occurs at some center and then this process repeats at another center \textit{etc}., which leads to the generation of a large noise. In contrast, in the regime of solid motion with friction above the second threshold $V_{\text{s}}$, the solid slides as a whole due to the over-barrier motion, and, therefore, the noise is suppressed. Thus, the physics of pinned periodic/elastic objects is relevant for the low-density state in a 2D electron system in silicon MOSFETs. These experimental results are also consistent with numerical simulations of the dynamics of a 2D electron system forming a Wigner solid in the presence of charged impurities \cite{reichhardt2001,reichhardt2016}.

\section{Summary}

In summary, we have reviewed the exciting properties of strongly interacting fermions in 2D electron and hole systems, in particular,  transport and thermodynamic data on the metallic side of the $B=0$ metal-insulator transition and recent advances in our understanding of the low-density state on the insulating side of the transition.  A complementary review \cite{qie2018} of the low-density state is available in the same volume.\\

\funding{AAS was supported by RFBR 18-02-00368 and 19-02-00196, RAS, and the Russian Government Contract. SVK was supported by NSF Grant No.\ 1309337 and BSF Grant No.\ 2012210.}

\acknowledgments{We are grateful to V.~T. Dolgopolov for careful reading of the manuscript and illuminating discussions.}

\conflictsofinterest{The authors declare no conflict of interest.}

\reftitle{References}

\end{document}